\documentclass[aps,reprint,floatfix,nofootinbib, superscriptaddress,longbibliography,prb]{revtex4-2}
\usepackage{natbib}
\usepackage{hyperref}
\usepackage[latin9]{inputenc}
\usepackage{amsmath}
\usepackage{amssymb}

\makeatletter

\usepackage{amsfonts}
\usepackage{xcolor}
\usepackage{graphicx}

\makeatother

\newcommand{\ooff}{$\mathrm{Rb}\mathrm{Eu}\mathrm{Fe}_{4}\mathrm{As}_{4}$}

\begin{document}

\title{Melting of vortex lattice in magnetic superconductor $\mathrm{Rb}\mathrm{Eu}\mathrm{Fe}_{4}\mathrm{As}_{4}$}

\author{A.\ E.\ Koshelev}
\affiliation{Materials Science Division, Argonne National Laboratory, 9700 South Cass Avenue, Lemont, IL 60439, USA}
\author{K.\ Willa}
\affiliation{Materials Science Division, Argonne National Laboratory, 9700 South Cass Avenue, Lemont, IL 60439, USA}
\affiliation{Institute for Solid-State Physics, Karlsruhe Institute of Technology, 76021 Karlsruhe, Germany}
\author{R.\ Willa}
\affiliation{Materials Science Division, Argonne National Laboratory, 9700 South Cass Avenue, Lemont, IL 60439, USA}
\affiliation{Institute for Theory of Condensed Matter, Karlsruhe Institute of Technology, 76131 Karlsruhe, Germany}
\author{M.\ P.\ Smylie}
\affiliation{Materials Science Division, Argonne National Laboratory, 9700 South Cass Avenue, Lemont, IL 60439, USA}
\affiliation{Department of Physics and Astronomy, Hofstra University, Hempstead, New York 11549}
\author{J.-K.\ Bao}
\affiliation{Materials Science Division, Argonne National Laboratory, 9700 South Cass Avenue, Lemont, IL 60439, USA}
\author{D.\ Y.\ Chung}
\affiliation{Materials Science Division, Argonne National Laboratory, 9700 South Cass Avenue, Lemont, IL 60439, USA}
\author{M.\ G.\ Kanatzidis}
\affiliation{Materials Science Division, Argonne National Laboratory, 9700 South Cass Avenue, Lemont, IL 60439, USA}
\affiliation{Department of Chemistry, Northwestern University, Evanston, Illinois, 60208, USA}
\author{W.-K.\ Kwok}
\affiliation{Materials Science Division, Argonne National Laboratory, 9700 South Cass Avenue, Lemont, IL 60439, USA}
\author{U.\ Welp}
\affiliation{Materials Science Division, Argonne National Laboratory, 9700 South Cass Avenue, Lemont, IL 60439, USA}

\date{\today }

\begin{abstract}
The iron-based superconductors are characterized by strong fluctuations due to high transition temperatures and small coherence lengths.
We investigate fluctuation behavior in the magnetic iron-pnictide superconductor {$\mathrm{Rb}\mathrm{Eu}\mathrm{Fe}_{4}\mathrm{As}_{4}$} by calorimetry and transport. We find that the broadening of the specific-heat transition in magnetic fields is very well described by the lowest-Landau-level scaling.  
We report calorimetric and transport observations for vortex-lattice melting, which is seen as a sharp drop of the resistivity and a step of the specific heat at the magnetic-field-dependent temperature. The melting line in the temperature/magnetic-field plane lies noticeably below the upper-critical-field line and its location is in quantitative agreement with theoretical predictions without fitting parameters. Finally, we compare the melting behavior of {$\mathrm{Rb}\mathrm{Eu}\mathrm{Fe}_{4}\mathrm{As}_{4}$} with other superconducting  materials showing that thermal fluctuations of vortices are not as prevalent as in the high-temperature superconducting cuprates, yet they still noticeably influence the properties of the  vortex  matter.
\end{abstract}
\maketitle

\section{Introduction}

Mean-field theory of type II superconductors predicts the formation of a periodic vortex lattice as a result of a continuous phase transition from the normal state at the upper critical field \cite{Abrikosov1957}. Thermal fluctuations qualitatively modify this scenario \cite{LarkinVarlBook2005}. The upper critical field becomes a crossover separating the normal phase from the vortex liquid state. This intermediate vortex liquid freezes into the lattice state via a first-order transition and the corresponding vortex-lattice melting field lies below the upper critical field.  

The extent of the vortex-liquid region varies widely between different superconducting materials and depends on the strength of thermal fluctuations, quantitatively characterized by the Ginzburg-Levanyuk number, $Gi$ \cite{Levanyuk1959, Ginzburg1960, LarkinVarlBook2005}. 
This number is uniquely determined by the superconducting parameters as
\begin{align}
	Gi=\frac{1}{2}\left(\frac{8\pi^2\lambda_{ab0}^2 T_{c} }{\Phi_0^2 \xi_{c0}}\right)^2.
	\label{eq:Gi}
\end{align}
Here and below $\xi_{ab0}$ and $\xi_{c0}$ are the coherence lengths and  $\lambda_{ab0}$ and $\lambda_{c0}$ are the London penetration depths%
\footnote{In quantitative estimates for $Gi$, we take the Ginzburg-Landau values for the coherence length and the London penetration depth rather than their low-temperature values.}.
At temperature $T \!-\! T_c \!\sim\! Gi\, T_c$ the fluctuation contribution to the specific heat is approximately equal to the specific-heat jump at the superconducting transition.
The magnetic field increases the fluctuation width of the transition. In the field range $B\!>\!Gi\,B_{c2}^{\prime}T_{c}$, the fluctuation broadening is determined by the field-dependent  Ginzburg-Levanyuk number\cite{LarkinVarlBook2005}, $Gi(B)\!=\! Gi^{1/3} \left[{B}/({B_{c2}^{\prime}T_{c})}\right]^{2/3}$ , where $B_{c2}^{\prime}\!=\!|dB_{c2}/dT|$ is the linear slope of the upper critical field at $T_c$.
The Ginzburg-Levanyuk numbers for several representative superconducting materials are summarized in Table \ref{Table:Gi}. In conventional type-II superconductors the Ginzburg-Levanyuk number is very small (e.g.\ for niobium $Gi\approx 7\cdot 10^{-12}$), and the vortex liquid occupies only a tiny and hard-to-detect strip below the upper critical field.

\begin{table*}
\label{Table:Gi} 
\begin{center}
\begin{tabular}{|c|c|c|c|c|c|c|}
\hline 
	\makebox[7em]{}
	& \makebox[6em]{$\mathrm{Nb}$}
	& \makebox[6em]{$\mathrm{Nb}_{3}\mathrm{Sn}$}
	& \makebox[6em]{$\mathrm{SnMo}_{6}\mathrm{S}_{8}$}
	& \makebox[6em]{$\mathrm{RbEuFe}_{4}\mathrm{As}_{4}$}
	& \makebox[6em]{$\mathrm{FeSe}$}
	& \makebox[6em]{$\mathrm{YBa}_{2}\mathrm{Cu}_{3}\mathrm{O}_{7}$}
	\tabularnewline
	\hline 
	Refs. & \cite{FinnemorePhysRev.149.231,DaSilvaPhysica69} & \cite{LortzPhysRevB.74.104502} & \cite{PetrovicPhysRevLett.103.257001} & \cite{Smylie2018c, Willa2018b} &
	\cite{Klein2019} & \cite{WelpPhysRevLett.76.4809,SchillingNature1996,*SchillingPhysRevLett.78.4833,*SchillingPhysRevB.58.11157}\tabularnewline
	\hline 
	\hline 
	$T_{c}${[}K{]} & 9.25 & 18 & 14.2 & 36.5 & 9.1 & 93.7\tabularnewline
	\hline 
	$B_{c2}^{\prime}${[}T/K{]} & 0.044 & 1.6 & 2.9 & 4.2 & 3.2 & 1.8\tabularnewline
	\hline 
	$\gamma$ & 1 & 1 & 1 & 1.7 & 4 & 7.8\tabularnewline
	\hline 
	$\Delta C$[kJ/K$\cdot$m$^{3}$] & 13 & 55 & 22 & 59 & 3.3 & 42\tabularnewline
	\hline 
	$\xi_{ab0}${[}nm{]} & 28.6 & 3.4 & 2.8 & 1.46 & 3.4 & 1.4\tabularnewline
	\hline 
	$\lambda_{ab0}${[}nm{]} & 21.3 & 61.7 & 133 & 98 & 357 & 75\tabularnewline
	\hline 
	$Gi$ & 6.9$\cdot$10$^{-12}$ & 1.3$\cdot$10$^{-7}$ & 2.5$\cdot$10$^{-6}$ & 5.3$\cdot$10$^{-5}$ & 6.1$\cdot$10$^{-4}$ & 2$\cdot$10$^{-3}$\tabularnewline
    \hline 
\end{tabular}

\caption {Superconducting parameters and the corresponding Ginzburg-Levanyuk numbers, Eq.~\eqref{eq:Gi}, for several materials. Here $\xi_{ab0}$ and $\lambda_{ab0}$ are the Ginzburg-Landau values of the coherence length and London penetration depth. The former is extracted from the linear slope of the upper critical field, $\xi_{ab0}=\sqrt{\Phi_{0}/2\pi T_{c}B_{c2}^{\prime}}$ 		with $B_{c2}^{\prime}\equiv dB_{c2,c}/dT$ for $T\rightarrow T_{c}$, while the latter is extracted from the specific heat jump, $\lambda_{ab0}=\Phi_{0}/\left(4\pi\xi_{ab0}\sqrt{2\pi\,\Delta\!C\,T_{c}}\right)$. $\gamma$ is the anisotropy of the upper critical field. }
\end{center}
\end{table*}

Soon after the discovery of the cuprate high-temperature superconductors, it became clear that the  Ginzburg-Levanyuk number in these materials is exceptionally large, $\sim\! 10^{-3}\text{-}10^{-2}$, due to their very high transition temperatures, small coherence lengths and large anisotropies. Early theoretical estimates based on the Lindemann criterion \cite{HoughtonPhysRevB.40.6763} suggested that 
the melting magnetic field is significantly lower than the upper critical field and
the vortex liquid state occupies a substantial portion of the temperature-field phase diagram.
The transport measurements performed on high-quality $\mathrm{Y}\mathrm{Ba}_{2}\mathrm{Cu}_{3}\mathrm{O}_{7-\delta}$ (YBCO) single crystals showed that the resistivity sharply drops to zero from a finite value at the field-dependent temperature $T_m(B)$ \cite{SafarPhysRevLett.69.824, KwokFWVDCM1992, CharalambousPhysRevB.45.5091}, which was interpreted as a manifestation of the first-order melting transition.  Later thermodynamic measurements have convincingly supported this interpretation: it was demonstrated that the magnetization features a small step  \cite{LiangPhysRevLett.76.835, WelpPhysRevLett.76.4809, NishizakiPhysRevB.53.82} while the specific heat has a peak and a step at the transition point \cite{SchillingNature1996, *SchillingPhysRevLett.78.4833, *SchillingPhysRevB.58.11157, RoulinScience1996, *RoulinJEW1996, *RoulinPhysRevLett.80.1722, *RevazPhysRevB.58.11153, BouquetNature2001}. 

The magnetic field at which melting occurs in YBCO is typically one quarter of the upper critical field, i.e., even though the melting is well separated from the mean-field crossover, it occurs at a comparable magnetic field strength. Consequently, the liquid and crystal state occupy comparable fractions of the vortex-matter phase space.  This situation is different in the much more anisotropic cuprate $\mathrm{Bi}_{2}\mathrm{Sr}_{2}\mathrm{Ca}\mathrm{Cu}_{2}\mathrm{O}_{8-\delta}$ (BSCCO), which represents an extreme scenario. 
The maximum melting field in this compound typically ranges from 300 G for optimal doping to 800 G in overdoped samples \cite{KhaykovichPhysRevLett.76.2555,OoiPhysC1998,BeidenkopfPhysRevLett.98.167004}. This is 
several thousand times smaller than the upper critical field ($\sim $ 100 -- 200 T) 
meaning that the liquid state occupies most of the vortex-matter region in the temperature-magnetic field phase diagram.
Such low melting fields allowed for direct monitoring of the crystalline long-range order across the transition by techniques sensitive to magnetic-field contrast, among which are neutron scattering \cite{CubittNat1993}, muon spin rotation \cite{LeePhysRevLett.71.3862}, and direct scanning Hall-probe imaging \cite{OralPhysRevLett.80.3610}.
In addition, due to its first-order character, the melting transition is accompanied by the noticeable magnetization jump \cite{PastorizaPhysRevLett.72.2951, ZeldovNature1995, *FuchsZMDTOK1996, FarrellPhysRevB.53.11807}. The observation of the first-order transition via the magnetization jump has been also reported for the compound $(\mathrm{La}_{1-x}\mathrm{Sr}_{x})_{2}\mathrm{Cu}\mathrm{O}_{4}$ (LSCO) \cite{SasagawaPhysRevLett.80.4297}. Depending on the doping level, the anisotropy factor of this compound varies between 20 and 50  situating this cuprate's fluctuation strength between that of YBCO and BSCCO. 

Extensive investigation of the vortex-lattice melting in high-temperature cuprates motivated more careful examination of this phenomenon in conventional superconductors. Specific-heat features associated with melting were reported for Nb$_3$Sn \cite{LortzPhysRevB.74.104502,LortzPhysRevB.75.094503} and the Chevrel phase SnMo$_6$S$_8$ \cite{PetrovicPhysRevLett.103.257001}. Even though these isotropic materials are characterized by high transition temperatures ($18$ K for Nb$_3$Sn and $14.2$ K for SnMo$_6$S$_8$) and upper critical fields ($\sim\! 29$ T for Nb$_3$Sn and $>\! 25$ T for SnMo$_6$S$_8$), the melting field still is located very close to the upper critical field and the vortex liquid occupies a very small fraction of the vortex-matter phase space. In general, observation of the melting transition in any material requires high-quality uniform single crystals. A small amount of disorder is sufficient to transform the first-order transition into the continuous glass transition. 

In terms of thermal fluctuations, iron pnictides are situated in between cuprates and conventional superconductors. In particular, the 122 and 1144 compounds---named after their chemical composition $A \mathrm{Fe}_{2}\mathrm{As}_{2}$ \cite{Paglione2010,Stewart2011,Hosono2015}, and $A B \mathrm{Fe}_{4}\mathrm{As}_{4}$ \cite{Iyo2016} respectively---have transition temperatures above $30$ K, $c$-axis upper critical fields reaching $80$ T, Ginzburg-Landau parameter $\kappa \!=\! \lambda_{ab}/\xi_{ab} \!\approx\! 70$, and a relatively small superconducting
anisotropy $1.5-2.5$ \cite{YuanNat2009, WelpPhysRevB.79.094505, SunPhysRevB.80.144515, TarantiniPhysRevB.84.184522}. These parameters combine to provide a Ginzburg-Levanyuk number $Gi\approx 5\cdot 10^{-5}-2\cdot10^{-4}$, of intermediate magnitude. 
One of the fluctuation effects observed in  122 compounds is a noticeable diamagnetic response above the superconducting transition in finite magnetic fields \cite{MosqueiraPhysRevB.83.094519, *Salem-SuguiPhysRevB.80.014518, *RamosAlvarezPhysRevB.92.094508}.   
The observation of the vortex-lattice melting has been reported in $\mathrm{Ba}_{0.5}\mathrm{K}_{0.5}\mathrm{Fe}_{2}\mathrm{As}_{2}$ \cite{MakPhysRevB.87.214523} using specific heat, thermal expansion, and magnetization measurements. In calorimetry, the transition has been seen as a small step in the temperature-dependent specific heat and a small peak develops at magnetic fields above $10$ T. In the temperature range $29-33$ K the melting field is approximately half of the upper critical field. 

The recently discovered magnetically ordered iron-pnictide superconductor {\ooff} \cite{Liu2016b, Bao2018, Smylie2018c, Willa2018b, Stolyarov2018, JacksonPhysRevB2018} has a transition temperature of $36.8$ K and slopes of the upper critical field 4.2 T/K and 7 T/K for the $c$-axis and $ab$-plane direction, respectively \cite{Smylie2018c, Willa2018b}. This compound also represents a rare case of magnetism coexisting with superconductivity. In fact, long-range magnetic order of $\mathrm{Eu}^{2+}$ magnetic moments develops below 15K without apparent suppression of superconductivity.
In this paper, we report resistive and calorimetric observations of the vortex-lattice melting in single crystals of this material. We show that the melting transition reveals itself through an abrupt drop of the resistivity and a step in the specific heat at the field-dependent temperature $T_m(B)$. The location of the melting line in the phase diagram is similar to that of Ba$_{0.5}$K$_{0.5}$Fe$_{2}$As$_{2}$. We compare the transition line with the theoretical predictions using experimental parameters of the material and find quantitative agreement without fitting parameters.

\section{Theoretical background}
\subsection{Lowest-Landau level scaling of specific heat}
The shape of the superconducting order parameter near the upper critical field within mean-field theory is given by the lowest-Landau level (LLL) wave function for a particle with charge $2e$ \cite{Abrikosov1957}.  
At high magnetic field $B\gg GiB_{c2}^{\prime}T_{c}$ the fluctuating order parameter near the mean-field transition can be well approximated as a linear combination of LLL wave functions, while contributions from higher Landau levels can be treated within Gaussian approximation. In this LLL regime, the fluctuation width of the transition is determined by the field-dependent Ginzburg number $Gi(B)$ and increases proportionally to $B^{2/3}$.      

Theoretical considerations  \cite{ThoulessPhysRevLett.34.946, BrezinPhysRevLett.65.1949, TesanovicPhysRevB.49.4064, LiRosenPhysRevB.70.144521, RosensteinRevModPhys.82.109} predict that within LLL regime, the superconducting contribution to the specific heat $C(B,T)\!-\!C_{n}(B,T)$ divided by the mean-field superconducting specific heat $C_{\mathrm{MF}}(T)$ has to be a universal function of a single scaling parameter
\begin{equation}
a_{T}=r_{sc}\frac{T-T_{c2}(B)}{\left(BT\right)^{2/3}}\label{eq:ScalingParam}
\end{equation}
with the coefficient
\begin{equation}
r_{sc}=\left(\frac{B_{c2}^{\prime}\sqrt{2T_{c}}}{\sqrt{Gi}}\right)^{2/3},\label{eq:ScCoef}
\end{equation}
and $T_{c2}(B)=T_c-B/B_{c2}^{\prime}$ the mean-field transition temperature at fixed magnetic field $B$.
This parameter corresponds to a shift of the temperature with respect to the mean-field transition normalized to the fluctuation broadening, $a_{T} \approx (T \!-\! T_{c2}(B))/Gi(B)T_{c2}(B)$. The scaling property implies that the experimental specific heat in the vicinity of the mean-field transition $T_{c2}(B)$ can be represented as
\begin{equation}
C(B,T)=C_{n}(B,T)+C_{\mathrm{MF}}(T)\;c_{sc}\!\left[r_{sc}\frac{T\!-\!T_{c2}(B)}{\left(BT\right)^{2/3}}\right].
\label{eq:CShape}
\end{equation}
The scaling function $c_{sc}(a_{T})$ in this equation captures the strongest temperature dependence near $T_{c2}(B)$, while all smooth contributions can be absorbed into the normal and superconducting backgrounds. Its precise shape follows from the corresponding free-energy scaling function that has been computed in Refs.~\cite{LiPhysRevB.65.220504,LiRosenPhysRevB.70.144521}.

\subsection{Vortex-lattice melting}
The dependence of the melting magnetic field of the vortex lattice on parameters of superconductors and temperature has been quantitatively investigated in two regimes.
At high fields, close to the upper critical field, $H_{c2}(T)$,  one can use the lowest Landau level  (LLL) approximation. In this regime, the melting transition has been studied by Monte-Carlo simulations \cite{SasikPhysRevLett.75.2582, *HuPhysRevB.56.2788} and analytically, using an elaborated evaluation of the free energies of the liquid and crystal states \cite{LiPhysRevB.65.220504, RosensteinRevModPhys.82.109}.  At low magnetic fields, in the London regime  [$B \!\ll\! H_{c2}(T)$], the vortex-lattice melting was investigated by Monte-Carlo simulations of the model of interacting lines \cite{NordborgPhysRevB.58.14556} and the uniformly-frustrated XY model \cite{ChenPhysRevB.55.11766, *KoshelevPhysRevB.56.11201}. It was also demonstrated that these two approaches give consistent results \cite{KoshelevPhysRevB.59.4358}.

The melting criterion in the LLL regime derived in Ref.\ \cite{LiPhysRevB.65.220504} is%
\footnote{This result also can be approximately expressed via the field dependent Ginzburg-Levanyuk number as $(T_{c2} \!-\! T_{m})/{T_{c2}} \!=\! 7.57\;Gi(B)$, where $T_{c2}$ and $T_{m}$ are the upper critical and melting temperatures at fixed field.}
\begin{align}
a_{T}=-9.5
\label{eq:MeltLLL}
\end{align}
where the parameter $a_{T}$ is defined by Eqs.~\eqref{eq:ScalingParam} and \eqref{eq:ScCoef}.  For easier comparison with the low-field regime, we rewrite the criterion \eqref{eq:MeltLLL} as
\begin{align}
\left(\frac{0.137\varepsilon_{0}\xi_{c}}{b_{m}T}\right)^{2}\left(1-b_{m}\right)^{3}=1\label{eq:MeltLLL-Comp}
\end{align}
with the $c$-axis coherence length $\xi_{c}(T)$, the reduced field $b_{m} \!=\! B_{m}(T)/H_{c2}(T)$, and the vortex line-energy scale $\varepsilon_{0} \!=\! [\Phi_{0}/4\pi\lambda_{ab}(T)]^{2}$. 
Note that although the LLL regime is only justified for $1-b_{m}\ll 1$, it usually works well beyond this range, i.e., at lower fields. Nevertheless, we point that this equation fails at describing the low-field behavior correctly, $b_{m}\ll 1$. In the London limit, $B\ll H_{c2}(T)$, the melting field rather reads $B_{m} \!\approx\! \Phi_{0}[{0.1\varepsilon_{0}}/({T\gamma})]^{2}$ \cite{KoshelevPhysRevB.59.4358} or
\begin{align}
b_{m}\approx\left(\frac{\varepsilon_{0}\xi_{c}}{4T}\right)^{2}.\label{eq:MeltLond}
\end{align}
Comparing this result with the LLL one in Eq.~\eqref{eq:MeltLLL-Comp}, we can introduce the interpolation formula
\begin{align}
\left[\frac{\varepsilon_{0}\xi_{c}\left(1-0.46b_{m}\right)}{4T}\right]^{2}\left(1-b_{m}\right)^{3}=b_{m}.
\label{eq:MeltInterp}
\end{align}
Assuming the mean-field temperature dependences, $\varepsilon_{0}=\varepsilon_{00}(1-t)$ and $\xi_{c}=\xi_{c0}(1-t)^{-1/2}$ with $t=T/T_c$, and the relation $Gi=(T_c/\xi_{c0}\varepsilon_{00})^2/8$ following from Eq.~\eqref{eq:Gi}, we can represent the above equation in the form
\begin{align}
\frac{\left(1-0.46b_m\right)\left(1-b_m\right)^{3/2}}{8\sqrt{2b_m}}=\frac{\sqrt{Gi}\,t}{\sqrt{1-t}}.
\label{eq:MeltInterpGi}
\end{align}
Here the parameter $t_{\mathrm{eff}} \equiv t \sqrt{Gi/(1-t)}$ on the right-hand side may be interpreted as an effective temperature, which determines the strength of thermal fluctuations. The solution of this equation provides the reduced melting field in the wide range $H_{c1}\ll B_m<H_{c2}$. It does not apply to describe the region of extremely low fields, where the distance between the vortex lines becomes comparable to $\lambda_{ab}(T)$. In addition, the mean-field approximation assumed in Eq.\ \eqref{eq:MeltInterpGi} breaks down in the region of strong fluctuations, $1-t \!<\! Gi$, where the critical 3D XY behavior develops. 

\section{Experimental details}

Single crystals of {\ooff} were grown in RbAs flux, as described in Ref.\ \cite{Bao2018}.
For specific heat measurements, we mounted small (100$\mu$m$\times$100$\mu$m$\times$10$\mu$m), uncut, platelet-shaped single crystals onto a SiN membrane based nanocalorimeter platform \cite{Tagliati2012, Willa2017} using Apiezon grease. The probe was then inserted into a three-axis vector magnet (1T-1T-9T), where we estimate the field to be aligned with the crystal axes up to $\pm$3 degrees. The orientation of the specific heat platform could be changed from perpendicular to parallel to the 9T $z$-axis of the magnet. The specific heat was obtained from $ac$ measurements ($f = 1\mathrm{Hz}$ and  $\delta T \sim 0.1\mathrm{K}$) and the signal was recorded with a Synktex lock-in amplifier. Resistivity samples were prepared by cutting larger {\ooff} crystals into bars and attaching 4 wires with Silver Epoxy. Then they were mounted into the same three-axis cryostat. The resistivity was then measured in a 4-point configuration at a constant current of 0.1mA. The data presented here is consistent with heat capacity and resistivity data published earlier in Refs.\ \cite{Smylie2018c, Willa2018b}.
\begin{figure}[ptb]
\includegraphics[width=0.4\textwidth]{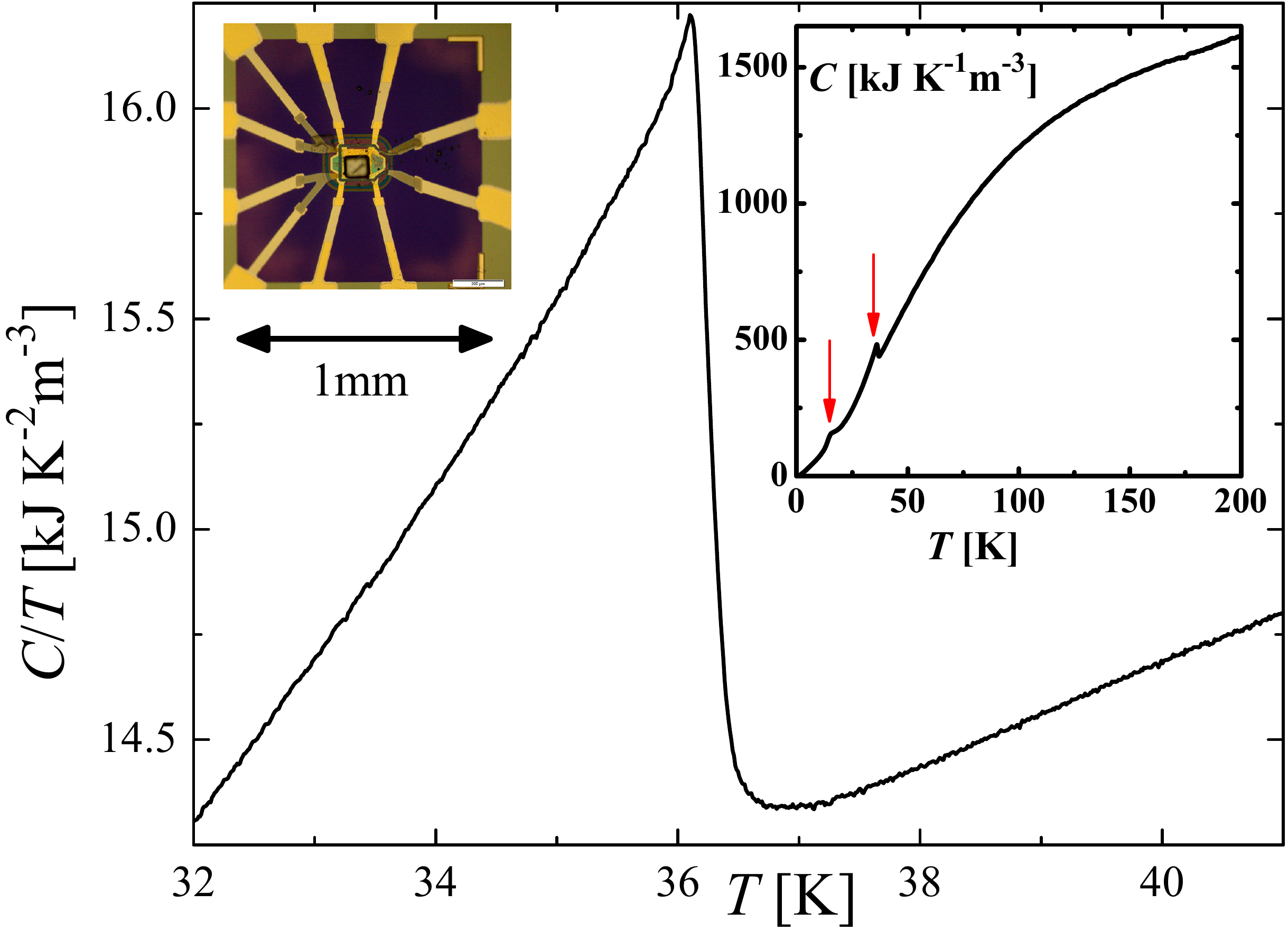}
\caption{Specific heat over temperature of a {\ooff} single crystal at the superconducting transition at 36.5K in zero field. The inset shows the specific heat during cooldown from high temperatures. The superconducting and the magnetic transitions are marked by the arrows. The picture in the upper left corner shows a microscope image of the measured crystal on the nanocalorimeter platform. }
\label{fig:transition}
\end{figure}
\begin{figure}[ptb]
\includegraphics[width=0.4\textwidth]{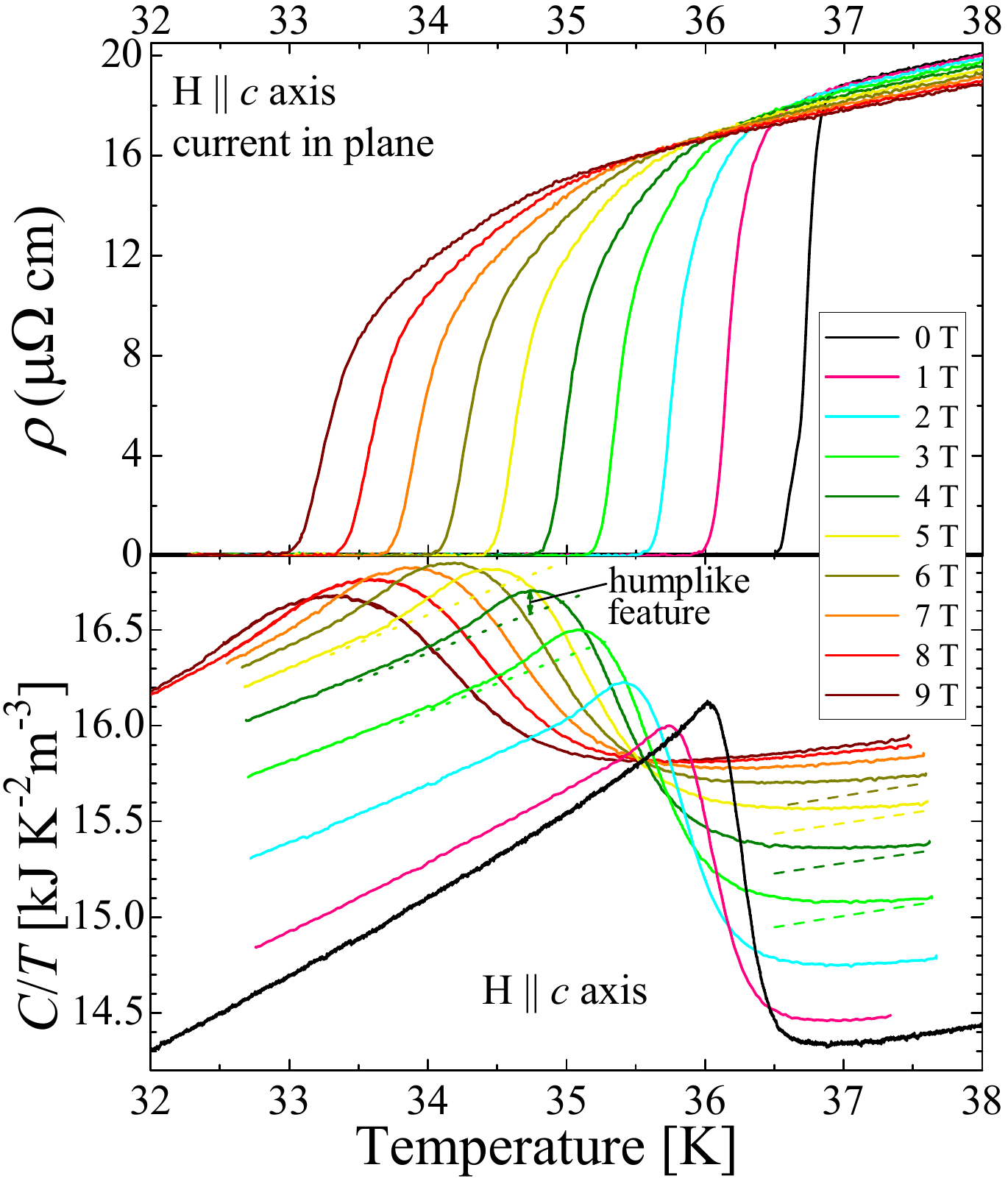}
\caption{
\textit{Top:} In plane resistivity for various magnetic fields along the $c$-axis for another {\ooff} single crystal. \textit{Bottom:} Superconducting transition of {\ooff} as seen in specific heat with $c$-axis magnetic fields up to 9T. The transition broadens and shifts to lower temperatures with increasing field. The hump visible on the low-temperature side of the transition is taken as an indication for vortex lattice melting. The dotted and dashed lines in the lower plot show the representative superconducting mean-field and normal backgrounds, respectively, extracted from the scaling analysis, see text.}	
\label{fig:vortex_melting}
\end{figure}

\section{Results and discussion}
\begin{figure}[b]	
\includegraphics[width=0.45\textwidth]{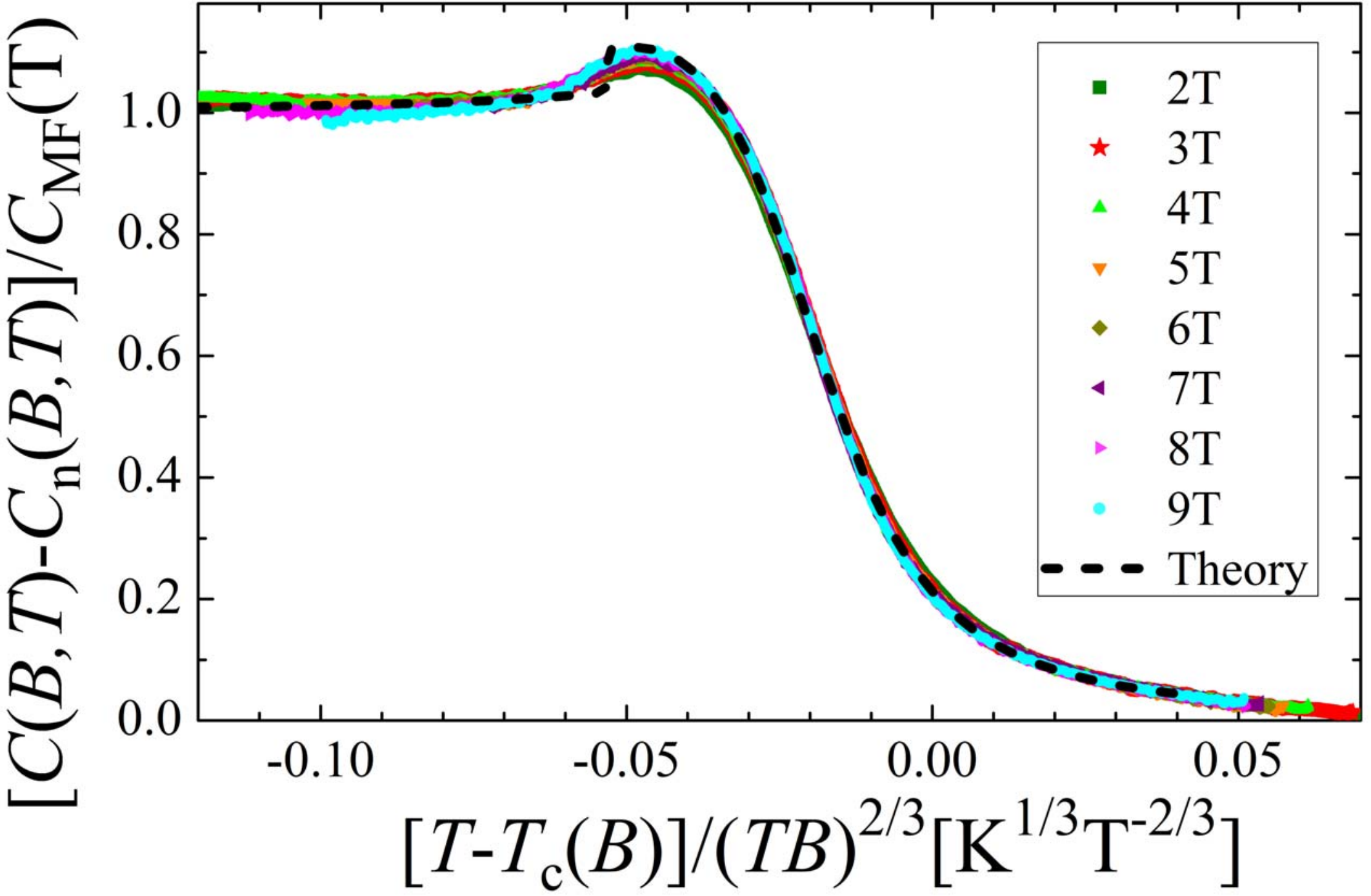}
\caption{Scaling plot of the specific heat  in the lowest-Landau level regime for high magnetic fields suggested by Eq.\ \eqref{eq:CShape}. The black dashed line is the theoretical LLL scaling function following from calculations in Refs.\ \cite{LiPhysRevB.65.220504,LiRosenPhysRevB.70.144521}.}
\label{fig:LLLScaling}
\end{figure}
\begin{figure}[t]		
\includegraphics[width=0.48\textwidth]{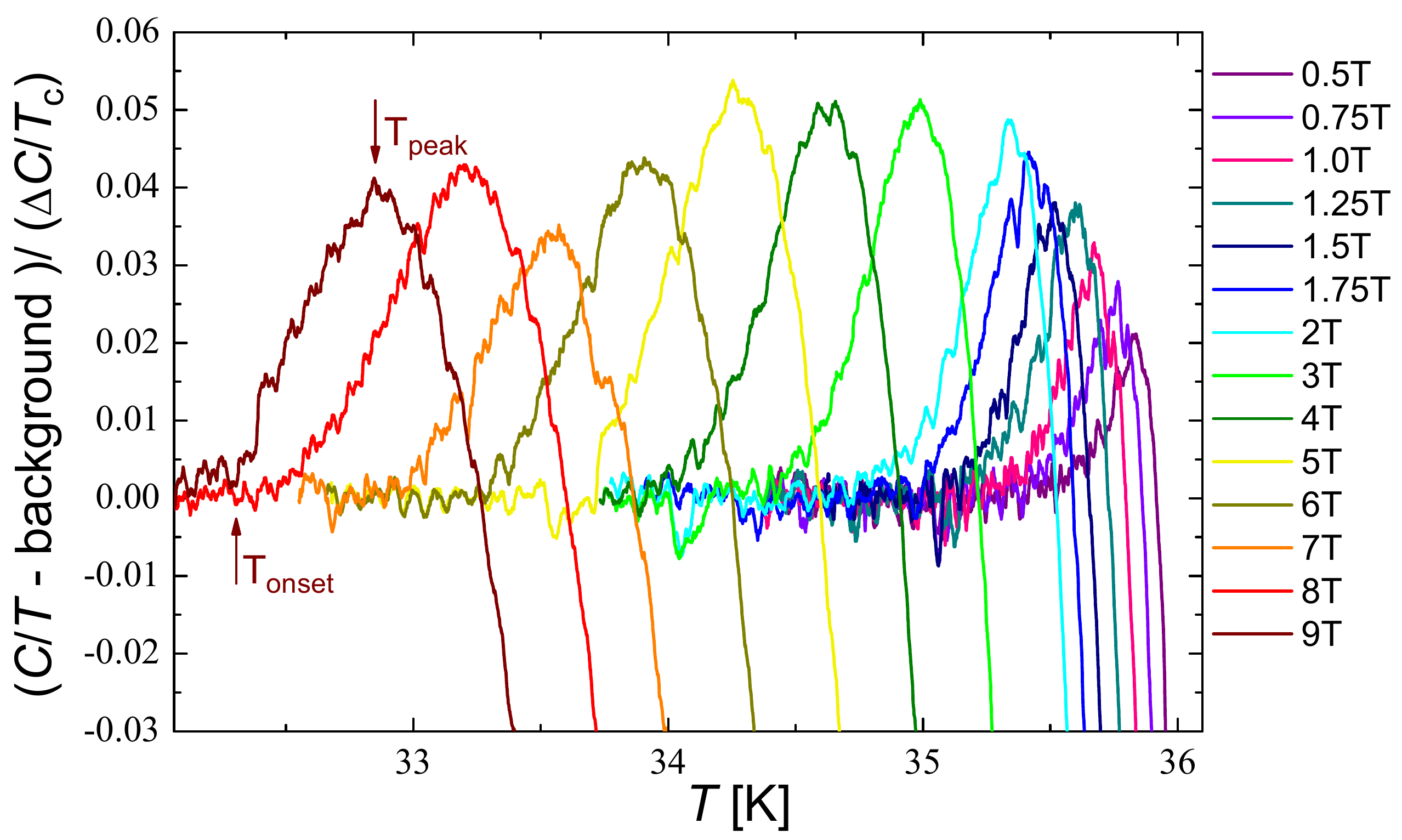}
\caption{Steplike feature in the specific heat associated with the vortex melting transition. The curves are obtained by subtracting linear fits for the low-temperature background contribution to enhance the visibility of the melting feature. The onset and peak positions of the step were extracted for the melting line. }
\label{fig:DCT-melt}
\end{figure}

The zero-field superconducting transition of a {\ooff} single crystal as seen in specific heat is shown in Fig.\ \ref{fig:transition}. The  specific-heat jump at the transition amounts to $\Delta C/T_c = 1.65 \pm 0.05$ $\mathrm{kJ/K}^2\mathrm{m}^3$ 
\footnote{In this paper, we use volume specific heat in units of $\mathrm{J}/\mathrm{K\; m}^3$, because it is this quantity that enters in all thermodynamic relations for superconductors. On the other hand, in experimental papers the molar specific heat in units of $\mathrm{J/K\; mol}$ is usually presented.  The conversion can be done using the molar volume which for {\ooff} is approximately equal to 120.5 $\mathrm{cm}^3/\mathrm{mol}$.}.
A noticeable upturn below the transition indicates that the broadening is caused by thermal fluctuations. The calorimetry data recorded at finite magnetic fields parallel to the $c$-axis, see Fig.\ \ref{fig:vortex_melting}(bottom), reveal a suppression of the transition temperature and a considerable broadening with increasing field strength. Additionally, a humplike feature below the transition temperature is a characteristic indicator of entropy excess associated with the vortex liquid state. The vertical shift of the $C/T$ curves are caused by the magnetic contribution to the specific heat arising from magnetic fluctuations of the $\mathrm{Eu}^{2+}$ moments, as discussed in Ref.\ \cite{Willa2018b}.

We find that the broadening of the specific-heat transition with the increasing magnetic field is in excellent agreement with theoretical predictions for an intrinsic fluctuation mechanism. Figure \ref{fig:LLLScaling} shows the LLL scaling plot of the specific heat in high magnetic fields, as suggested by Eq.\ \eqref{eq:CShape}. 
The theoretical scaling function shown in the figure is based on calculations reported in Refs.\ \cite{LiPhysRevB.65.220504,LiRosenPhysRevB.70.144521}.\footnote{The dominating contribution to the specific-heat scaling function is given by the last term in Eq.\ (57) of Ref.\ \cite{LiRosenPhysRevB.70.144521}, which is proportional to the second derivative of the free-energy scaling function. The shape of this function plotted in Fig.\ \ref{fig:LLLScaling} has been provided to us by the authors of that paper.}
At the first stage, we approximated both $C_n(B,T)$ and $C_{\mathrm{MF}}(T)$ by linear functions of the temperature and used also $r_{sc}$ and $T_{c2}(B)$ in Eq.\ \eqref{eq:ScalingParam} as fit parameters for each magnetic field. At the second stage, we fixed the fit parameters of $C_{\mathrm{MF}}(T)$ and $r_{sc}$ at averaged values for fields from 3 to 9 T, where they change weakly. The best match is achieved for $r_{sc}\approx 177$ K$^{-1/3}$T$^{2/3}$ and
$C_{\mathrm{MF}}(T)\approx -183.6+6.845 T[\mathrm{K}]$ kJ/K$\cdot$m$^3$ for $31\,\mathrm{K}<T<35\,\mathrm{K}$. The representative superconducting mean-field and normal backgrounds extracted from this analysis are shown in Fig.\ \ref{fig:vortex_melting}. As one can see, the procedure leads to the almost perfect collapse of the data to the theoretical scaling curve. We note, however, that the estimate of the parameter $r_{sc}$ from Eq.\ \eqref{eq:ScCoef} gives somewhat higher value $\approx 290$ K$^{-1/3}$T$^{2/3}$. The step in the theoretical curve 
where it features a small discontinuity of $\approx\! 10\%$
is due to the first-order vortex lattice melting transition, which we discuss below. 

\begin{figure}[t]
\includegraphics[width=0.48\textwidth]{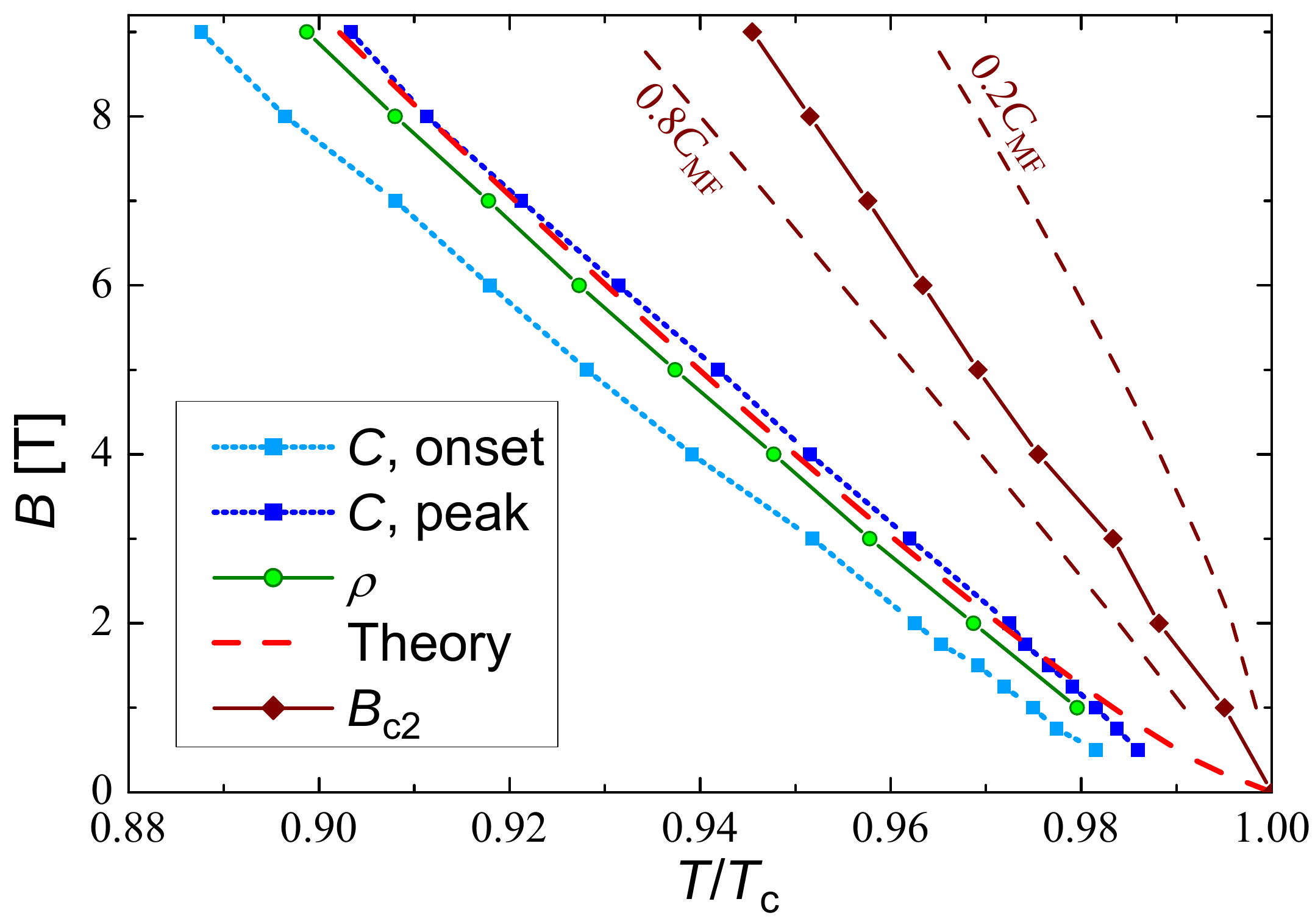}
\caption{The vortex melting lines extracted from the specific heat, see Fig.\ \ref{fig:DCT-melt}, and  from vanishing resistivity together with the theoretical curve calculated from Eq.~\eqref{eq:MeltInterp}. For comparison, we also show the plot of the upper critical field extracted from the specific-heat data. %
The dashed lines illustrate the fluctuation broadening of the transition and are obtained using the criteria 
$[C(B,T)\!-\!C_n(B,T)]/C_{\mathrm{MF}}\!=\!0.2$ and $0.8$.}
\label{fig:PhaseDiagr}
\end{figure}

Transport measurements show that the resistive transition broadens considerably at finite magnetic fields,  see Fig.\ \ref{fig:vortex_melting}(top). The transition somewhat sharpens when resistivity drops below $\sim60\%$ of the normal level forming a kinklike feature. We interpret this feature as an indication of the first-order-melting resistivity jump smeared by inhomogeneities in the crystal.
Similar behavior has been observed in moderately-disordered YBCO samples, while very uniform YBCO crystals feature a very sharp jump. We set the threshold resistance 0.2 $\mu \Omega\cdot$cm as the criterion for determining the melting temperature in transport measurements. In specific heat, the contribution of the vortex-lattice melting is determined by subtracting from the raw data the linear contribution below the excess-entropy hump. The result, scaled to the jump height at $T_c$ in zero field, $\Delta C/T_c$, is shown in Fig.~\ref{fig:DCT-melt}. The relative height of the hump, 4-5\% of $\Delta C/T_c$, is in agreement with theoretical expectations for vortex-lattice melting \cite{LiPhysRevB.65.220504}.
The spike corresponding to the first-order transition is not visible, probably due to a small number of impurities in the sample. 
This behavior corresponds to the melting transition broadened by inhomogeneities and it is very similar to 
Nb$_3$Sn \cite{LortzPhysRevB.74.104502,LortzPhysRevB.75.094503}, SnMo$_6$S$_8$ \cite{PetrovicPhysRevLett.103.257001}, and $\mathrm{Ba}_{0.5}\mathrm{K}_{0.5}\mathrm{Fe}_{2}\mathrm{As}_{2}$ \cite{MakPhysRevB.87.214523}. In contrast, the specific-heat data in YBCO at high magnetic fields show closer-to-ideal behavior with a pronounced spike at the transition and a distinct step \cite{SchillingNature1996,*SchillingPhysRevLett.78.4833,*SchillingPhysRevB.58.11157}. This difference is not only due to higher homogeneity of the YBCO crystals but also due to much larger Ginzburg-Levanyuk number of this material leading to the large separation between $B_m$ and $B_{c2}$.
Bounds for the melting transition temperature are extracted by evaluating the onset and peak positions of the specific-heat hump, as marked in Fig.\ \ref{fig:DCT-melt}. The clean-limit melting temperature will lie between these two values. The phase diagram in  Fig.\ \ref{fig:PhaseDiagr} shows the melting line extracted from resistivity and calorimetry data, together with the $H_{c2}$-line extracted from the specific-heat curves by the entropy conserving construction. We also show the theoretical curve obtained using Eq.~\eqref{eq:MeltInterp} with the Ginzburg-Levanyuk parameters extracted from the experimental $H_{c2}$ slopes and specific-heat jump at $T_{c}$, $\xi_{ab0}=1.46$ nm, $\lambda_{ab0}=98$ nm, and $\gamma=1.7$. The melting transition line is in excellent agreement with theoretical expectations. We emphasize that the theoretical curve is parameter-free and results purely from the experimental data. We also point out that the location of the melting line is very close to that of a similar compound Ba$_{0.5}$K$_{0.5}$Fe$_{2}$As$_{2}$ \cite{MakPhysRevB.87.214523}.

\begin{figure}[tb]
\includegraphics[width=0.48\textwidth]{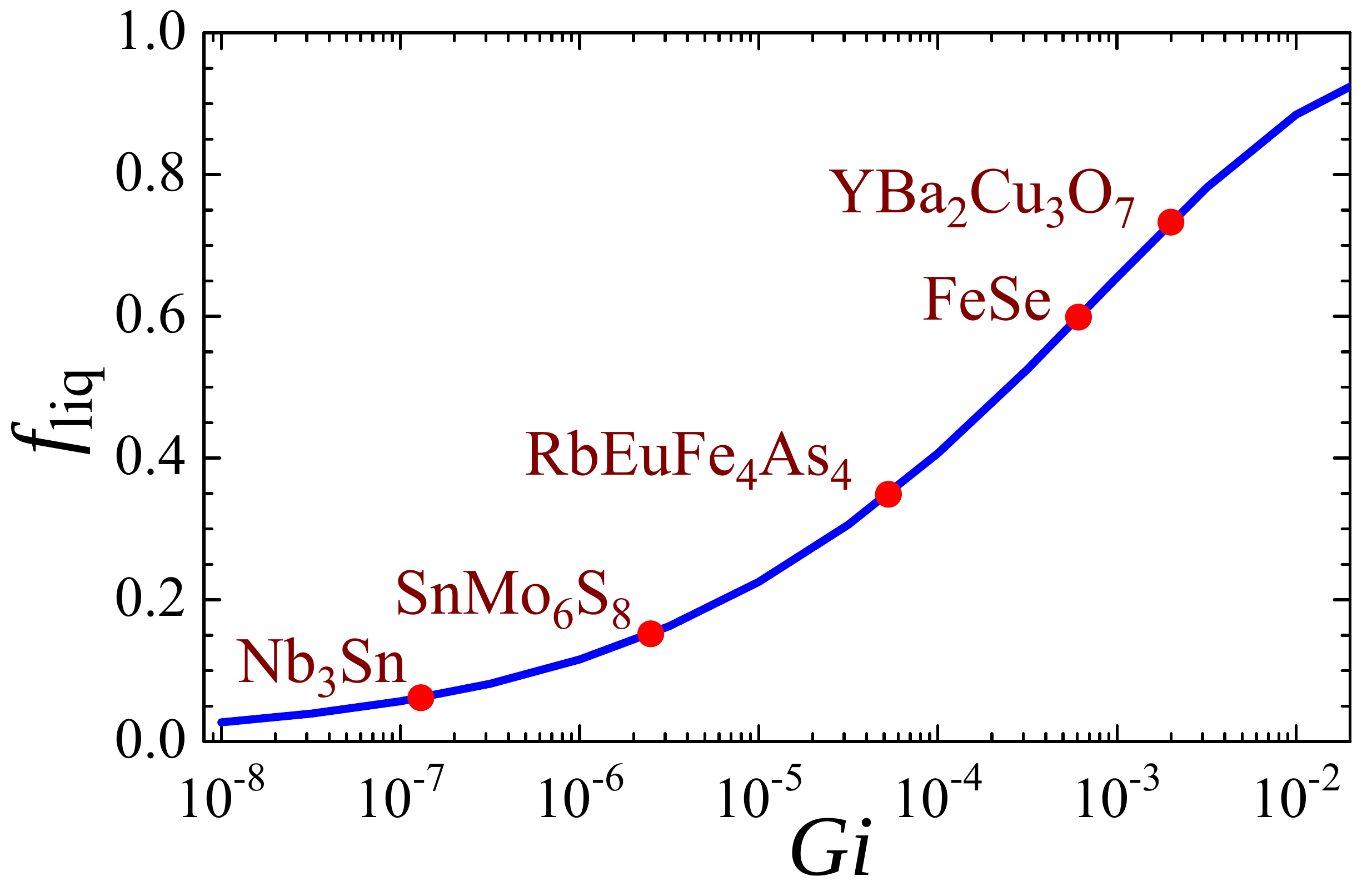}
\caption{Dependence of the area fraction occupied by the liquid state, $f_{\mathrm{liq}}$ within the GL vortex-matter region ($0.75T_c<T<T_c$, $B<B_{c2}(T)$) on the Ginzburg-Levanyuk number.}
\label{fig:Fliq}
\end{figure}

To place our results within a broader context, we compare the melting behavior of {\ooff} with other 
superconducting materials. Even though the Ginzburg-Levanyuk number, Eq.~\eqref{eq:Gi}, is the  standard parameter for characterizing thermal fluctuations, it does not directly convey the relative relevance of the themodynamic phases (liquid and solid) of the vortex matter. Here, we introduce a more intuitive parameter $f_{\mathrm{liq}}$ quantifying the fraction of liquid phase in the vortex-matter phase diagram within the Ginzburg-Landau (GL) regime, $T_{\mathrm{GL}}<T<T_{c}$. Here we arbitrarily define the boundary of GL regime as $T_{\mathrm{GL}}=3T_{c}/4$.
Explicitly, this parameter is 
\begin{align}
f_{\mathrm{liq}} \equiv \frac{\int_{T_{\mathrm{GL}}}^{T_{c}}(H_{c2}\!-\!B_{m})dT}{\int_{T_{\mathrm{GL}}}^{T_{c}}H_{c2}dT}\!=32\int\limits_{3/4}^{1}(1-t)(1-b_m)dt.
\end{align}
We use the interpolation formula in Eq.~\eqref{eq:MeltInterpGi} for its quantitative evaluation. 
Figure \ref{fig:Fliq} shows the computed dependence  $f_{\mathrm{liq}}(Gi)$ with points for four representative materials. For weakly-fluctuating materials, in which the melting  falls inside the LLL regime, we obtain $f_{\mathrm{liq}}\approx 12.8Gi^{1/3}\ll 1$. 
The liquid phase occupies more than half of the phase space in the GL region already for very small Ginzburg-Levanyuk numbers, $Gi>6\cdot10^{-5}$. {\ooff} is only slightly below this threshold and its location clearly indicates that, as other iron-based superconductors, this material is in between conventional superconductors and cuprates.

\begin{figure}[t]
\includegraphics[width=0.45\textwidth]{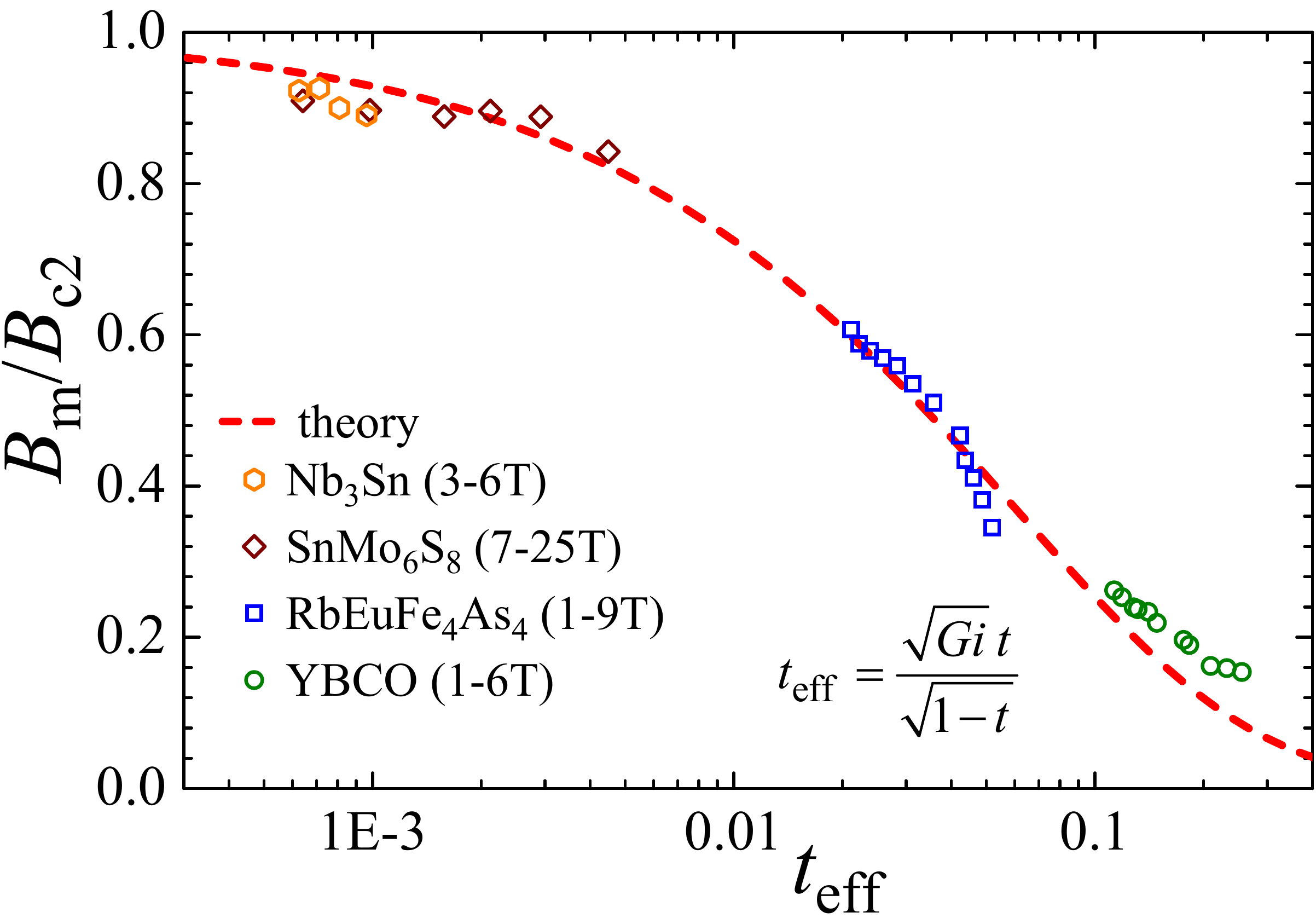}
\caption{Universal plot for the vortex-lattice melting transition. The theoretical curve is determined by Eq.~\eqref{eq:MeltInterpGi}. The references for used material's data are listed in Table \ref{Table:Gi}.}
\label{fig:UnivMelt}
\end{figure}

As follows immediately from the form of Eq.~\eqref{eq:MeltInterpGi}, the reduced melting field $B_m/B_{c2}$ in the mean-field GL region is a universal function of the effective reduced temperature parameter $t_{\mathrm{eff}}=\sqrt{Gi}\, t/\sqrt{1\!-\!t}$ which determines the strength of thermal fluctuations. Figure \ref{fig:UnivMelt} shows this universal melting plot computed from Eq.~\eqref{eq:MeltInterpGi} together with the points extracted from experimental data for four different materials, Nb$_{3}$Sn \cite{LortzPhysRevB.74.104502}, SnMo$_{6}$S$_{8}$ \cite{PetrovicPhysRevLett.103.257001}, RbEuFe$_{4}$As$_{4}$ (this work), YBa$_{2}$Cu$_{3}$O$_{7}$ \cite{WelpPhysRevLett.76.4809,SchillingNature1996,*SchillingPhysRevLett.78.4833,*SchillingPhysRevB.58.11157}. We can see that the experimental melting lines follow the universal trend suggested by theory.  

\section{Summary}
We investigate superconducting thermal fluctuations in single crystals of the magnetic iron-pnictide superconductor {\ooff} via transport and calorimetric measurements. We find that the broadening of the specific-heat transition in the magnetic-field range 2-9 T is well described by the lowest-Landau-level scaling.   
We observe the vortex-lattice melting transition, which is manifested as a steep drop of the resistivity and a step of the specific heat at the field-dependent temperature $T_m(B)$.  Melting takes place considerably below the upper critical field. The location of the melting line in the temperature-magnetic field plane is in quantitative agreement with theoretical predictions without fitting parameters.  We demonstrate that the reduced melting fields for different materials follow a universal dependence on the effective temperature parameter which determines the strength of thermal fluctuations. Our observations imply that even though thermal fluctuations of vortices are not as prevalent as in HTSC cuprates, they still  noticeably influence the properties of the vortex matter. In particular, we estimate that the liquid phase occupies roughly 40\% of the vortex-matter space within the Ginzburg-Landau region.

\begin{acknowledgments}
We would like to thank Dingping Li and Baruch Rosenstein for discussion of the scaling behavior of the specific heat in the LLL regime and providing to us the theoretical scaling function.
The work was supported by the US Department of Energy, Office of Science, Basic Energy Sciences, Materials Sciences and Engineering Division. K. W. and R. W. acknowledge support from the Swiss National Science Foundation through the Postdoc Mobility program. 
\end{acknowledgments}
\vfill
\pagebreak
\bibliography{VLmelt}
\end{document}